\documentclass[]{article}

\usepackage[margin=1in]{geometry}
\usepackage[utf8]{inputenc}

\usepackage{algorithm}
\usepackage[noend]{algorithmic}
\usepackage{amssymb}
\usepackage{booktabs}
\usepackage{changes}
\usepackage{comment}
\usepackage{graphicx}
\usepackage{mathtools}
\usepackage{url}
\usepackage{multirow}
\usepackage{array}
\usepackage{esvect}
\usepackage{float}
\usepackage{nicefrac}
\usepackage{rotating}
\usepackage{lscape}
\usepackage{outlines}
\usepackage[nodisplayskipstretch]{setspace}
\usepackage{csvsimple}
\usepackage{longtable}
\usepackage{subcaption}

\usepackage{scalerel}
\usepackage{tikz}
\usetikzlibrary{svg.path}

\definecolor{orcidlogocol}{HTML}{A6CE39}
\tikzset{
  orcidlogo/.pic={
    \fill[orcidlogocol] svg{M256,128c0,70.7-57.3,128-128,128C57.3,256,0,198.7,0,128C0,57.3,57.3,0,128,0C198.7,0,256,57.3,256,128z};
    \fill[white] svg{M86.3,186.2H70.9V79.1h15.4v48.4V186.2z}
                 svg{M108.9,79.1h41.6c39.6,0,57,28.3,57,53.6c0,27.5-21.5,53.6-56.8,53.6h-41.8V79.1z M124.3,172.4h24.5c34.9,0,42.9-26.5,42.9-39.7c0-21.5-13.7-39.7-43.7-39.7h-23.7V172.4z}
                 svg{M88.7,56.8c0,5.5-4.5,10.1-10.1,10.1c-5.6,0-10.1-4.6-10.1-10.1c0-5.6,4.5-10.1,10.1-10.1C84.2,46.7,88.7,51.3,88.7,56.8z};
  }
}

\newcommand\orcidicon[1]{\href{https://orcid.org/#1}{\mbox{\scalerel*{
\begin{tikzpicture}[yscale=-1,transform shape]
\pic{orcidlogo};
\end{tikzpicture}
}{|}}}}

\usepackage{hyperref} 

\floatname{algorithm}{Procedure}

\usepackage[space]{grffile}

\definechangesauthor[color=red]{is}
\definechangesauthor[color=blue]{js}
\definechangesauthor[color=green]{rs}

\usepackage[]{todonotes}

\newcommand{\link}{\url{https://sybrandt.com/2019/partition}}

\newif\ifarxiv
\arxivtrue

\title{Hypergraph Partitioning With Embeddings}

\author{
    Justin Sybrandt \\
    Clemson University \\
    jsybran@clemson.edu
    \and
    Ruslan Shaydulin \\
    Clemson University \\
    rshaydu@clemson.edu
    \and
    Ilya~Safro\\
    Clemson University \\
    isafro@clemson.edu
}

\date{}

\title{Hypergraph Partitioning with Embeddings}

\begin{document}

\maketitle
\begin{abstract}

  Problems in scientific computing, such as distributing large sparse matrix
  operations, have analogous formulations as hypergraph partitioning problems. A
  hypergraph is a generalization of a traditional graph wherein ``hyperedges''
  may connect any number of nodes. As a result, hypergraph partitioning is an
  NP-Hard problem to both solve or approximate.  State-of-the-art algorithms
  that solve this problem follow the multilevel paradigm, which begins by
  iteratively ``coarsening'' the input hypergraph to smaller problem instances
  that share key structural features. Once identifying an approximate problem
  that is small enough to be solved directly, that solution can be interpolated
  and refined to the original problem. While this strategy represents an
  excellent trade off between quality and running time, it is sensitive to
  coarsening strategy. In this work we propose using graph embeddings of the
  initial hypergraph in order to ensure that coarsened problem instances retrain
  key structural features. Our approach prioritizes coarsening within self-similar
  regions within the input graph, and leads to significantly improved solution
  quality across a range of considered hypergraphs.
  \\ {\bf Reproducibility:}
  All source code, plots and experimental data are available at \link{}.

\end{abstract}

\ifarxiv
\section{Introduction}
\label{sec:introduction}
\fi

\ifarxiv
Hypergraphs
\fi
provide the formalism needed to solve problems
consisting of interconnected item sets.  Similar to a traditional graph, the
hypergraph has the added generalization that ``hyperedges'' may connect any
number of nodes.  Domains such as very-large-scale integration for creating
integrated circuits~\cite{karypis1999multilevel}, machine
learning~\cite{zhou2007learning,hein2013total,zhang2017re}, parallel
algorithms~\cite{catalyurek1999hypergraph}, combinatorial scientific
computing~\cite{hendrickson2006combinatorial}, and social network
analysis~\cite{shepherd1990transient,zhang2010hypergraph} all contain
significant and challenging instances of hypergraph problems.  One important
problem, \emph{Hypergraph partitioning}, involves dividing the nodes of a
hypergraph among $k$ similarly-sized disjoint sets while reducing the number of
hyperedges that span multiple partitions.  In the context of load balancing,
this is the problem of dividing logical threads (nodes) that share data
dependencies (hyperedges) among available machines (partitions) in order to
balance the number of threads per machine and minimize communication overhead.
However, hypergraph partitioning is both NP-Hard to
solve~\cite{lengauer2012combinatorial} and approximate~\cite{bui1992finding}.

\begin{figure}
  \centering
  \includegraphics[width=0.7\linewidth]{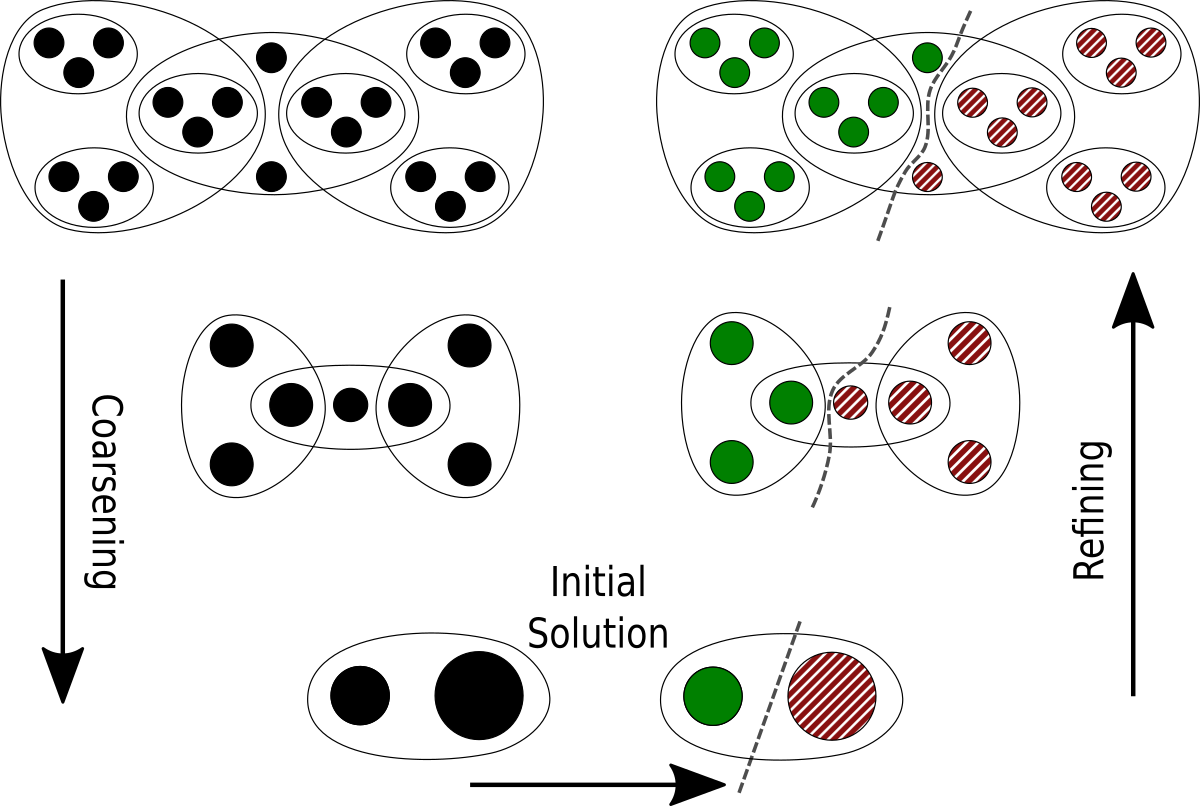}
  \caption{A standard V-cycle, consisting of coarsening, and initial partition,
  and uncoarsening.}
  \label{fig:v_cycle}
\end{figure}

Therefore, state-of-the-art partitioners apply heuristically-backed algorithms
to overcome these inherent computational limitations~\cite{bulucc2016recent}. The
most common and effective technique is the \emph{multilevel
paradigm}~\cite{andre2018memetic, shaydulin2019relaxation,
karypis1999multilevel, boman2009advances, devine2006parallel}. The multilevel paradigm is well known to be successful beyond the  (hyper)graph partitioning in such areas as the cut-based problems on graphs~\cite{safro2009multilevel} and machine learning~\cite{sadrfaridpour2019engineering}. Multilevel
partitioners consist of three phases, referred to collectively as the
\emph{V-Cycle}: coarsening, the initial solution, and uncoarsening. We depict
these phases in Figure~\ref{fig:v_cycle}. The overarching idea behind this
technique is to find a problem instance that shares key structural
features with the input hypergraph, but is small enough to be partitioned
directly. The initial solution to this small analogous problem can then be gradually 
interpolated and refined to apply to the input hypergraph.

The small analogous problem is identified through an iterative \emph{coarsening}
process consisting of many levels. At each level, groups of similar nodes are
identified, and each is ``contracted'' into a single merged node at the next
more-coarse level.  While grouping nodes, the goal is to identify 
self-similar regions of the current hypergraph so that the more coarse problem
instances retain key structural features.  Most commonly, these coarsening
groups are formed by pairing nodes due to a similarity
measure~\cite{devine2006parallel, shaydulin2019relaxation} that heuristically or rigorously suggests to place both nodes in the same partition. An $n$-level
algorithm is one that identifies only one pair of coarsening partners at each
level~\cite{shhmss2016alenex}, while a $\log n$-level algorithm pairs many 
nodes each time~\cite{devine2006parallel}.  Coarsening stops once identifying a
sufficiently small hypergraph based on some criterion that indicates that this problem is possible to (almost) optimally solve on given computational resources. The \emph{initial
solution} can then be identified directly using the best available algorithm. Now, the solution is \emph{uncoarsened} back through the
levels in order to identify a solution to the original problem. Uncoarsening
consists of three sub-phases: expansion, interpolation, and refinement.
Expansion undoes the coarsening at a given level by ``expanding'' the current
level's coarsened nodes with those contracted in the prior. Next, interpolation
assigns each expanded node the partition label assigned to their corresponding
coarse representation.  Then, local refinement cheaply updates the partition
labels among the expanded nodes in order to improve the overall solution quality
for the next level. This process is repeated from the initial solution through
all coarsening levels and back to the original hypergraph, which final refinement solution is accepted as
the solution to the partitioning problem.

Because the strategy used to contract nodes determines the coarsening at each
level, the quality of the initial solution, and the behavior of interpolation
and refinement during uncoarsening, we find that this single factor can
dramatically effect partitioning quality. Other works exploring coarsening
strategies, such as relaxation-based~\cite{shaydulin2019relaxation} or
community-aware~\cite{hs2017sea} coarsening, arrive with a similar conclusion.

\subsection{Our Contribution}

\begin{figure}
  \centering
  \includegraphics[width=0.7\linewidth]{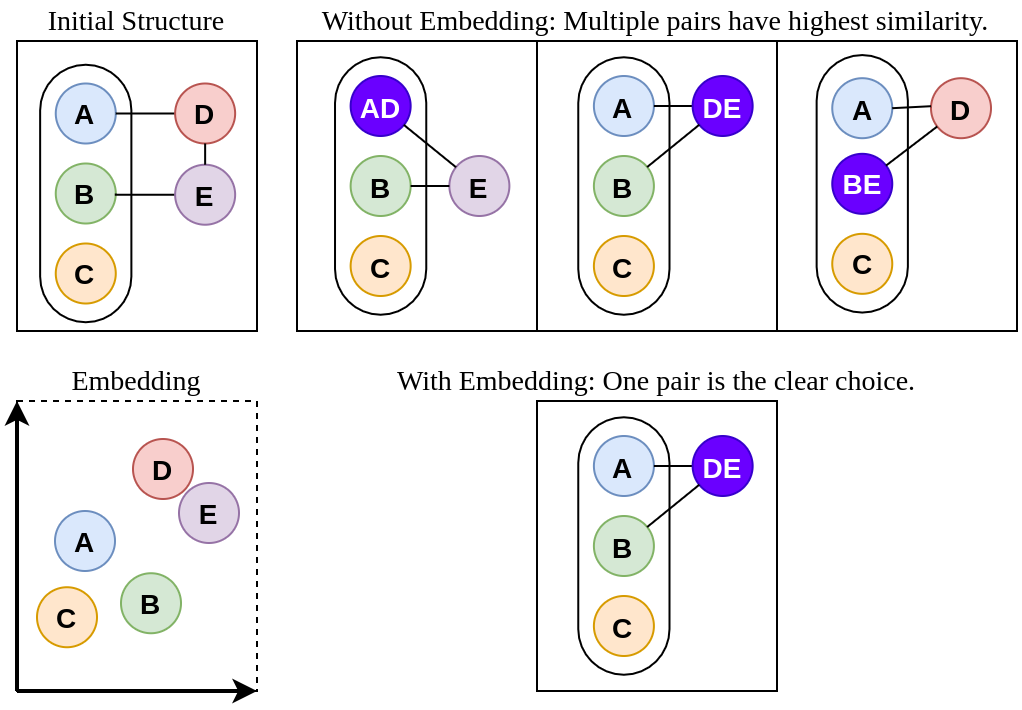}
  \caption{
    Typical heavy-edge coarsening results in three possible contractions of equal weight. Embedding-based coarsening breaks the tie for the pair that preserves global structure --- a cluster of weight 3 and of weight 2.
  }
  \label{fig:part_example}
\end{figure}

 We propose \emph{embedding-based coarsening}, a novel coarsening strategy that
leverages graph embeddings to prioritize the contraction of 
self-similar regions of the input hypergraph in order to retain global
structural features.  This approach augments the existing strategy that
contracts nodes based on their co-participation in small hyperedges by adding an
embedding-based term that can break ties among similarly ranked coarsening
pairs. A toy example of this phenomena is depicted in
Figure~\ref{fig:part_example}, wherein three potential coarsening pairs are
equally ranked by the traditional scheme, but embedding-based signals favor
the pair that retains both key clusters.

The field of graph embedding is evolving rapidly, and the proposed
embedding-based coarsening is designed to be agnostic with respect to any
particular technique, provided that similarities between nodes are encoded via
the dot product of embedding vectors. Specifically, our proposed technique
accepts a precomputed embedding as an auxiliary input per-hypergraph, and we
demonstrate that a wide range of existing embedding techniques improve
partitioning performance similarly. Decoupling partitioning from embedding
enables embedding-based coarsening to more easily benefit from future advances
in machine learning techniques. In order to apply embedding techniques designed
for classical graphs, we need a classical representation of each input
hypergraph. The star-expansion~\cite{agarwal2006higher} represents a hypergraph
as an undirected bipartite graph wherein hyperedges from the original structure
form a new layer of nodes. An edge between two nodes $i$ and $j$ in the
bipartite structure indicates that node $i$ participated in hyperedge $j$ within
the original structure. As opposed to other classical representations like the
clique-expansion, the star-expansion retains all relevant hypergraph
information, and is scalable for large graphs~\cite{agarwal2006higher}.
Furthermore, existing embedding techniques specifically designed for bipartite
graphs~\cite{sybrandt2019heterogeneous} apply to star-expanded graphs directly.

Given an input hypergraph and an embedding for each node of the input
structure, embedding-based coarsening follows this outline at each coarsening
level. First, each node is assigned a score equal to the highest dot product
between its embedding and each of its neighbors. Nodes are visited in decreasing
order by score. A visited node is matched from among its neighbors based on the
product of their classical edge-wise score, and the dot product of each node's
embedding. After matching nodes, based on whether we are performing $n$- or
$\log n$-level coarsening, mated nodes are contracted. Newly coarsened nodes are
assigned an embedding equal to the average embedding of all initial embeddings
contained within the coarse representation.

We implement our proposed coarsening strategy in both
KaHyPar~\cite{shhmss2016alenex}, which is a $n$-level partitioner with
state-of-the-art solution quality, as well as Zoltan~\cite{devine2006parallel},
which is a parallel $\log n$-level partitioner with high quality and
state-of-the-art speed. Furthermore, we compare the effect of various different
embedding techniques, including Node2Vec~\cite{grover2016node2vec},
Metapath2Vec++~\cite{dong2017metapath2vec}, and
FOBE/HOBE~\cite{sybrandt2019heterogeneous}, which were designed specifically for
bipartite graphs. We additionally compare the effect
of each embedding-based coarsening strategy with
hMetis~\cite{karypis1998hmetis},
Zoltan~\cite{devine2006parallel}, PaToH~\cite{ccatalyurek2011patoh}, KaHyPar
(with community-based coarsening~\cite{hs2017sea}), and KaHyPar Flow (with both
community-based coarsening and flow-based
refinement~\cite{heuer2018network})\footnote{
  Neither hMetis nor PaToH provide source code. Instead, we can only use
  pre-compiled binaries for comparison purposes.
}. We compare performance of each partitioner across 96 hypergraph from the
SuiteSparse Matrix Collection~\cite{davis2011university}. For each graph, we
compute one embedding using each of the proposed techniques to serve as an
auxiliary input across all trial. We compare quality across both the ``cut''
and ``connectivity'' objectives as well as for partition counts from 2 to 128
for each partitioner.
For each combination of experimental parameters we run 
20 trials in
order to compare the variance and establish significance with respect to
different random seeds. Overall, we produce over 500,000 individual trials.

We find that embedding-based coarsening has a significant improvement over the
state of the art that is especially pronounced for smaller partition counts
($<32$). In some cases, this leads to a solution quality that is improved by as
much as $400\%$. Because embedding-based coarsening replaces the traditionally
random visit order with one that prioritizes self-similar regions of the
hypergraph, we also observe an improvement in the standard deviation of quality. 
All experimental code, data, visualization
scripts, and a database of all experimental results, including all
hyperparameters per-trial, can be found at: \link{}.
A longer-form version of this work can be found online\footnote{\url{https://arxiv.org/abs/1909.04016}}.

\section{Notation and Preliminary Concepts}
\label{sec:prelim}

A hypergraph $H=(V,E)$ consists of nodes $v\in V$ and hyperedges $e\in E$. As
opposed to a traditional graph, each hyperedge may contain any non-empty subset
of $V$. The hypergraph partitioning problem is to divide $V$ into $k$ disjoint
subsets of similar size while minimizing a given objective function. Two common
objectives considered here are ``cut'' and ``connectivity.'' Cut measures the
number of hyperedges spanning more than one partition. If $\lambda(e)$ is the
number of partitions spanned by edge $e$, then the cut objective is defined as:
  ${|e\in E, \text{ such that } \lambda(e) > 1|}$.

The ``connectivity'' objective, also commonly referred to as ``$k-1$,''
penalizes each edge by the number of spanned partitions, namely, ${\sum_{e\in E} (\lambda(e) - 1})$. In the case of
$k=2$, this is equivalent to cut. 

\noindent{\bf Weights.}
Although we consider unweighted input hypergraphs (all nodes and hyperedges
count the same towards their corresponding objectives and constraints), the multilevel paradigm introduces
weights to intermediate level hypergraphs. Each node and hyperedge has a
corresponding weight ($w_v$ and $w_e$) equal to one for the input hypergraph.
During coarsening, if two nodes $v_i$ and $v_j$ are contracted into a new coarse
node $v'$, then $w_{v'}=w_{v_i} + w_{v_j}$. This new node $v'$ will also be
added to all edges originally containing either $v_i$ or $v_j$, before removing
those original nodes from the resulting coarse hypergraph. 
If during the coarsening two edges will contain the same subset of nodes ($e_1 = e_2$), they will be replaced with a new edge $e'$ containing the same nodes but with added weights: $w_{e'} = w_{e_1} + w_{e_2}$.
When solving for cut and connectivity for intermediate sub-problems 
during the
multilevel strategy, we introduce these weights into the objective:
\begin{equation}
  \text{weighted cut} = \sum_{e \in E, \lambda(e) > 1} w_e
\end{equation}
\begin{equation}
  \text{weighted connectivity} = \sum_{e\in E} (\lambda(e) - 1) w_e
\end{equation}

One issue during coarsening is the potential for individual nodes to accumulate a
disproportionate amount of weight. When this occurs, balanced partitioning can
become impossible at the coarsest level, especially if one node's weight exceeds
$|V|/k$. In order to avoid this negative effect, multilevel partitioners enforce
a weight tolerance $w^{(T)}$, which is parameterized by the user. No coarsening
partners may be contracted if their resulting coarse node would exceed this
limit.

\noindent{\bf Imbalance Constraint.} It is important to balance the number of nodes in
the resulting partitions.  Therefore, partitioners include an optimization
constraint to determine how uneven the resulting partitions are allowed to be.
For each partition $V_i \subset V$, given a predefined imbalance tolerance
$\alpha$, this constraint is defined as:
\begin{equation}
  \sum_{v' \in V_i} w_{v'} \leq
  (1+\alpha) \left\lceil \frac{1}{k} \sum_{v \in V} w_v\right\rceil
\end{equation}

\noindent{\bf Embeddings.} We use the function
$\epsilon:V\rightarrow\mathbb{R}^n$ to denote a pre-trained embedding.
Conceptually, this is a lookup table that assigns a node in the input hypergraph
to a real-valued $n$-dimensional vector. In our experiments we select $n=100$.
Note, higher-dimensional embeddings show no statistically significant performance difference in our experiments.

\section{Background and Related Work}
\label{sec:background}

\noindent{\bf Multilevel Partitioning.}
\ifarxiv
First introduced to speed up existing algorithms~\cite{barnard1994fast} and
inspired by multigrid and multiscale optimization strategies~\cite{vlsicad}, the
multilevel method  was quickly recognized as an effective approach to improve the
quality of (hyper)graph partitioning~\cite{karypis1998fast}, and is currently
considered to be one of the state-of-the-art methods for this
problem~\cite{bulucc2016recent}.  As introduced in
Section~\ref{sec:introduction}, the basic version of the multilevel paradigm solves problems by
following the V-cycle pattern that consists of coarsening, the initial solution,
and uncoarsening. This approach is effective because coarse hypergraphs are
easier to solve yet they retain global structural features of the original.
Coarse hypergraphs are created by iteratively merging multiple nodes at the
current ``finer'' level into single nodes at the ``coarser'' level.
Once sufficiently small, a partitioner can directly solve the coarsest problem instance using an algorithm that would normally be infeasible for large problems. The uncoarsening process then applies that solution through iteratively finer problem instances by expanding contracted nodes, interpolating coarse solutions onto finer problem instances, and refining intermediate solutions at each level. Once the uncoarsening process reaches the most-fine problem instance, the refined solution is accepted as the partitioning of the input hypergraph.

\fi

Usually, at each level of the coarsening process almost all nodes
have at least one merging partner, resulting in $\log n$ levels. This is the
approach used by Mondriaan~\cite{vastenhouw2005two},
hMetis2~\cite{karypis1999multilevel}, Zoltan~\cite{devine2006parallel}, and
PaToH~\cite{ccatalyurek2011patoh}.  However, KaHyPar~\cite{shhmss2016alenex}
implements an $n$-level approach where at each level only one pair of nodes is
contracted. 
\ifarxiv
The multilevel paradigm is the current gold-standard for hypergraph
partitioning, having achieved an excellent trade off between time and quality.
Unsurprisingly, most practical and state-of-the-art partitioners follow this
paradigm, including all methods considered in this work.  For an extensive
review of (hyper)graph partitioning methods, we refer the
reader to~\cite{bulucc2016recent,bichot2011graph}.
\fi

\noindent{\bf Coarsening Strategies.}
\ifarxiv
Multilevel partitioners follow heuristic strategies to identify groups of
nodes to contract during coarsening.
\fi
A good coarsening strategy is one that
groups together nodes that will ultimately share the same partition label,
meaning that the coarser solution can be interpolated to the finer solution
without a loss of quality. In practice, this loss of quality is to be expected,
which is why local refinement is common during uncoarsening. However, if
global structural features are not preserved, the loss of quality during
interpolation cannot be rectified through the fast local refinement process.
Therefore, the choice of coarsening heuristic is paramount.

Most heuristics used to identify nodes for contraction do so by scoring node
pairs, and most partitioners,
including Mondriaan~\cite{vastenhouw2005two},
hMetis2~\cite{karypis1999multilevel} and Zoltan~\cite{devine2006parallel},
measure the edge-wise inner product, or some variation.
The edge-wise inner-product is the Euclidean inner product of the weighted
hyperedge incidence vectors~\cite{devine2006parallel}. Edge weights are defined
formally in Section~\ref{sec:prelim}. Specifically, if $w_e$ is the weight of
hyperedge $e$, then the edge-wise inner product of nodes $u$ and $v$ is defined
as:
\begin{equation*}
  \sum_{u,v \in e \in E}{w_e}
\end{equation*}
Although this approach is simplistic, it is also very computationally inexpensive
and has provided a firm baseline. As mentioned, many variations exist, such as
\emph{absorption}, implemented in PaToH~\cite{ccatalyurek2011patoh}, and \emph{heavy
edge}, implemented in  hMetis2~\cite{karypis1999multilevel},
Parkway~\cite{trifunovic2008parallel}, and KaHyPar~\cite{hs2017sea}, as well as
a number of other normalization techniques, often based on node or hyperedge
degree. Heavy edge, which is of particular interest due to its simple
formulation and high performance, simply normalizes hyperedge weight by the expected
degree of the resulting hyperedge following contraction. If $|e|$ is the number
of nodes present in hyperedge $e$, then this score is:
\begin{equation*}
  S_E(u, v) = \sum_{u,v \in e \in E}\frac{w_e}{|e|-1}
  \label{eq:kahypar_inner_product}
\end{equation*}

One key limitation to the edge-wise score heuristics is that each only considers
local information around each node. Therefore, global structural features can be
collapsed during coarsening. This work seeks to use graph embeddings to provide
this global information, however prior work has attempted to provide similar
signals in alternate ways.
Shaydulin et al.
introduce \emph{algebraic distance} for hypergraphs, a relaxation-based
similarity measure that extends a similar approach from traditional
graphs~\cite{chen2011algebraic,ron2011relaxation}. This measure treats nodes as entities in a
mutually-reinforcing environment, which enables this technique to apply a fast
relaxation-based approach to supply a coordinate per node. Conceptually this
acts as a one-dimensional embedding, wherein two nodes receive a similar
coordinate if their neighborhoods are similar. This similarity measure is used
to quantify node similarities and assign weights to hyperedges.

Another approach to incorporate global information is \emph{community-aware}
coarsening, which uses clustering information to restrict matching between
communities. This approach, which is implemented in KaHyPar, makes the
assumption that nodes belonging to different clusters of the input hypergraph
should never be contracted. The proposed clustering is performed by a fast
global modularity-maximizing algorithm, leveraging the connection between
partitioning and clustering.  This modularity-based clustering, which groups
star-expanded nodes within a bipartite representation of a hypergraph,
identifies communities are internally dense and externally
sparse~\cite{newman2010networks}, which is desired for a good partitioning.  We
note, and discuss further in Section~\ref{sec:results}, that the clusters found
by this modularity-maximizing approach are similar to the self-similar regions within a
graph embedding.  However, in some scenarios the hard restriction to never merge
nodes across communities appears to be too strong. Instead, embedding-based
coarsening simply penalizes the contraction of nodes across clusters, allowing
more flexible decisions for nodes along the periphery.

\noindent{\bf Refinement.} While this work proposes a new coarsening strategy,
important work also explores the refinement stage of uncoarsening, wherein each
partitioner performs local search in order to improve the interpolated coarser
solution on the finer level. The typical strategy is the \emph{node-moving
heuristic}, wherein each expanded node at the newly refined level is given the
option of switching partition label.  A majority of hypergraph partitioners use
a variation of Fiduccia-Mattheyses~\cite{fiduccia1988linear} or
Kernighan-Lin~\cite{kernighan1970efficient} to perform these local
searches~\cite{heuer2018network, vastenhouw2005two, karypis1999multilevel,
devine2006parallel, ccatalyurek2011patoh, trifunovic2008parallel}.  Recently,
Heuer et al. introduced a flow-based refinement scheme for $k$-way hypergraph
partitioning~\cite{heuer2018network}, extending similar approaches from graph
partitioning~\cite{sanders2011engineering}.  This flow-based refinement, which
is implemented in KaHyPar, is considered as a temperate case within our
benchmark. As a result, we can compare the performance of embedding-based
coarsening without flow-based regiment, and vice-versa.

\subsection{Related Partitioning Strategies.}
There are a few coarsening
and partitioning strategies that are not included in our benchmark, but are
worth additional discussion.  \emph{Memetic partitioning}, also proposed for
KaHyPar, uses the principles of genetic algorithms to discover improved
partitioning solutions~\cite{andre2018memetic}.  This approach creates high
quality partitions by iterating through different ``generations'' of solutions,
starting with an initial generation produced by  KaHyPar run multiple times with
different seeds.  From the initial set, multiple combination operators ``breed''
new solutions by combining some number of ``parents'' to form new solutions.
Each iteration is designed to improve the population's average connectivity
metric. Combination operators are specifically posed such that offspring
solutions perform at least as good as its corresponding parents. While this
approach is demonstrated to improve overall hypergraph partitioning quality, it
does so by adding a meta process to the set of initial hypergraph solutions. We
anticipate that adding embedding-based coarsening as a method for generating a
high quality initial solution population may be a complimentary way to improve
the overall process.  \emph{Aggregative coarsening}~\cite{shaydulin2018sea} uses
ideas from algebraic multigrid. 
At each step of the coarsening
process a set of seed vertices is selected. Each seed then becomes a center of
an aggregate, with non-seeds assigned to seeds using different aggregation
rules. An aggregate at finer level forms a vertex at coarser level. Two
aggregation rules, based on inner product matching and stable matching were
explored.  Our embedding-based coarsening can be used within the aggregative
coarsening to inform the aggregation rules.

\subsection{Graph Embeddings}
\label{sec:background:hypergraph_embeddings}

\ifarxiv
Our embedding-based coarsening accepts embeddings for each node of the
input hypergraph as an auxiliary input. While we make the assumption that node
similarity is encoded through the dot product of embeddings, we do not depend on
any particular embedding technique. However, there are a range of embedding
methods that we consider in our benchmark due to their scalability and
applicability to the bipartite graphs produced by the star-expansion process.
At a high level, graph embeddings assign a real-valued vector of fixed size to
each node (and sometimes each edge) of an input graph. Therefore, techniques
such as non-negative matrix factorization, principal component analysis, or even
algebraic distance~\cite{shaydulin2019relaxation} can all apply as embeddings
from the perspective of embedding-based coarsening. While our early experiments
explored all of these and more, we found the most significant improvements when
using neural-network-based embeddings.

In all of the considered embedding techniques, various node
features are encoded via the dot product of node embeddings. Furthermore, new
techniques are published frequently that identify new ways to encode latent node
footers.  Rather than depend on a particular graph embedding technique, this
work simply assumes that some measure of global graph structure is encoded via
the dot product of embeddings, meaning that two nodes with a higher dot product
of embeddings will be more similar. In this manner, new advances in graph
embedding, or fine-tuned versions of existing algorithms for particular graphs,
can be introduced into our proposed strategy. Importantly, this work does not
seek to establish any graph embedding technique as inherently better for
hypergraph partitioning. \emph{Instead, we find that all considered embeddings greatly
improve solution quality}.
\fi

\noindent{\bf Neural Graph Embedding.}
The Deepwalk graph embedding~\cite{perozzi2014deepwalk}, which applies the
skip-gram model~\cite{mikolov2013distributed} to random walks of nodes, marks
the beginning of neural network graph embeddings. The node2vec
approach~\cite{grover2016node2vec} modifies Deepwalk to parameterize random walk
behavior, allowing walks to explore local regions or broad swaths of a graph. In
doing so, Grover et al. identify that node2vec graph embeddings can encode both
homophilic and structural latent features. Tsitsulin et al.  generalize the
formalism across a range of random-walk based graph embedding techniques, noting
that community-based, role-based, and structural features of nodes can all be
encoded in a single unified framework~\cite{tsitsulin2018verse}.

\noindent{\bf Bipartite Embeddings.}
Sybrandt et al.~\cite{sybrandt2019heterogeneous} explore a number of bipartite graph embedding
techniques when presenting First- and Higher-Order Bipartite Embedding (FOBE and
HOBE), including BiNE~\cite{gao2018bine}, and
Metapath2Vec++~\cite{dong2017metapath2vec}.
Due to the star expansion, bipartite embedding techniques are likely among the best suited for hypergraphs.
Therefore, we select a subset of bipartite embedding methods that performed the best in this prior work to explore here.
\ifarxiv
Bipartite Network Embeddings (BiNE)
generates walks that are weighted by a network centrality measure, and uses
these walks to fit both explicit and implicit relationships simultaneously.
However, this method performs similar to random embeddings for a range of link
prediction tasks in~\cite{sybrandt2019heterogeneous}, and for this reason is
omitted from the benchmark in this work. Metapath2Vec++ extends the Deepwalk
framework in two ways. First, random walks are restricted to follow particular
patterns of nodes. Second, node embeddings are computed using different
parameters based on that node's type. In the case of bipartite graphs, the only
random walk pattern is that of alternating node types, but the added flexibility
gained by parameterizing each side of a bipartite graph separately leads to
improved performance in~\cite{sybrandt2019heterogeneous}.
\fi

\noindent{\bf FOBE and HOBE.}
Because most readers will not be familiar with FOBE and HOBE, and because we
find these techniques are often the most useful for embedding-based coarsening, we
summarize these techniques here. Both methods create a set of observations, which are used to train embeddings.

Formally, if $G=(V,E)$ is a bipartite graph, $\Gamma(x)$ is the neighborhood of
node $x$, and $u,v\in V$ are nodes, then FOBE assigns a sampled score for the
$u,v$ pair as follows:
\begin{equation*}
  \mathbb{S}^{(\text{FOBE})}(u, v) =
  \begin{cases}
    1 & \Gamma(u) \cap \Gamma(v) \neq \emptyset \\
    1 & uv \in E \\
    0 & \text{otherwise}
  \end{cases}
\end{equation*}
These observations are fit by minimizing the following for each observed  $u,v$ pair, where $\epsilon(x)$ indicates the learned
embedding of node $x$:
\begin{equation*}
  \begin{aligned}
    \mathcal{L}^{(\text{FOBE})}(u, v) &=
      \sigma(\epsilon(u), \epsilon(v))
      \log\left(
        \frac{\mathbb{S}^{(\text{FOBE})}(u,v)}
             {\sigma(\epsilon(u),\epsilon(v))}
      \right),\\
    \text{where } \sigma(x) &=  \frac{1}{1+e^{-x}}.
  \end{aligned}
\end{equation*}

HOBE, in contrast, learns higher-ordered relationships that are weighted using
algebraic distance, the same underlying technique used
within relaxation-based coarsening~\cite{shaydulin2019relaxation}, which places nodes on the unit interval such that similar nodes receive similar coordinates.
Formally, the algebraic coordinate of node $u$ is determined
by this iterative process:
\begin{equation*}
  {\bf a}_{i+1}(u) =
  \frac{1}{2}\left(
    {\bf a}_{i}(u)
    + \frac{
      \sum\limits_{v\in\Gamma(u)} {\bf a}_{i}(v)|\Gamma(v)|^{-1}
    }{
      \sum\limits_{v\in\Gamma(u)} |\Gamma(v)|^{-1}
    }
  \right)
\end{equation*}
where ${\bf a}_0$ is randomly initialized, and the algebraic coordinate for $u$
is determined after a fixed number of steps $t$.
HOBE runs $R=10$ random restarts and summarizes the algebraid similarity between two nodes $u$ and $v$ in the following way:
\begin{equation*}
  \begin{aligned}
    s(u, v) &= \frac{\sqrt{R} - d(u, v)}{\sqrt{R}} \\
    \text{where } d(u, v) &=
    \sqrt{\sum_{r=1}^R\left({\bf a}_t^{(r)}(u) - {\bf a}_t^{(r)}(v)\right)^2}
  \end{aligned}
\end{equation*}

HOBE uses algebraic similarities to assign weights to node pairs by identifying highly similar shared neighbors:
following manner:
\begin{equation*}
  \begin{aligned}
    \mathbb{S}^{(\text{HOBE})}(u, v) &=
  \begin{cases}
    \alpha(u,v) & \Gamma(u)\cap\Gamma(v)\neq\emptyset \\
    \max\left(\begin{aligned}
      \max\limits_{x\in\Gamma(v)}\alpha(u, x),\\
      \max\limits_{x\in\Gamma(u)}\alpha(x, v)
    \end{aligned}\right)
    & uv \in E \\
    0 & \text{ otherwise }\\
  \end{cases}\\
  \text{where } \alpha(u, v) &=
    \max\limits_{x\in\Gamma(u)\cap\Gamma(v)}
    \min\left(
        s(u, x),
        s(v, x)
    \right)
  \end{aligned}
\end{equation*}

Embeddings are fit by minimizing the following loss for each $u$, $v$ pair:
\begin{equation*}
  \mathcal{L}^{(\text{HOBE})}(u, v) =
  \left(
  \mathbb{S}^{(\text{HOBE})}(u, v)
  - \max(0, \epsilon(u)^\intercal\epsilon(v))
  \right)^2
\end{equation*}

\noindent{\bf Combination Embeddings.}
To demonstrate the ability for embedding-based coarsening to apply to any given
embeddings, we explore the combination approach also presented by Sybrandt et
al. in~\cite{sybrandt2019heterogeneous}. This method learns a joint
representation for each node given multiple pre-trained embeddings. This
technique does not rely on any random walk strategy, and instead learns a
unified embedding per-node given the edge list as a set of embeddings per node.
The particular model combines a link-prediction objective with an auto-encoding
objective, and in doing so ensures that the resulting joint embedding captures
relevant structural signals that are needed to reproduce both the edge list as
well as the input embeddings. This technique is very similar to that presented
by Wang et al.~\cite{wang2016structural} in that it consists of two connected
auto encoders. The result of this method is an embedding that merges the
structural features present in a range of embeddings while preserving any useful
distinct features from across the set. We direct the reader
to~\cite{sybrandt2019heterogeneous} to find the specifics of this approach.

\noindent{\bf Deep Learning Graph Embedding.}
In addition to the above techniques, which are generally fast, scalable, and
parallel sizable, there are another set of deep-learning embedding techniques that
apply larger models to the problem of graph embedding. One popular technique,
the graph convolutional network~\cite{kipf2016semi}, constructs a neural network
in the same structure as the input graph, and embeddings are derived by a
``message-passing'' function that distributes node features among neighborhoods.
Another technique by Cao et al. learns deep representation by first constructing
a large co-occurrence matrix from a process of ``random-surfing'' following by
deep auto encoders~\cite{cao2016deep}. A similar auto-encoder-based approach is
presented by Wang et al.~\cite{wang2016structural}, wherein a pair of deep
auto-encoders both encode nodes independently, as well as ensure that similar
nodes are assigned similar embedding. While these deep-learning techniques do
achieve high quality results for relatively small graphs, these techniques are
less scalable than the previously discussed class of algorithms, due to their
larger model structure and the accompanying need for more graph samples.  While
these techniques could certainly improve the quality of embedding-based
coarsening for some hypergraphs, we designed our proposed technique to be
independent of any particular embedding, and evaluated our technique over a
large collection of hypergraphs and scenarios. As a result, the analysis of
deep-learning graph embedding techniques was infeasible for this work.

\section{Embedding-Based Coarsening}
\label{sec:method}

Embedding-based coarsening begins with a user-supplied hypergraph as well as an
embedding of each node. For instance, we use the
star-expansion~\cite{agarwal2006higher} of the hypergraph in order to apply a
range of embedding techniques designed for classical graphs. During coarsening,
nodes are visited in an order determined by the embeddings of each node's
neighborhood.  When visited, an unmatched node is paired with whichever neighbor
maximizes a combined measure of edge-wise inner product as well as embedding dot
product. After identifying matches, paired nodes are contracted into new coarse
nodes, which are assigned an embedding equal the average of all the initial embeddings it represents.
Because embedding-based coarsening preserves more global
structural features than other methods, the initial partitioning solution is
more applicable to the large-scale graph, resulting in higher partitioning
quality. We implement embedding-based coarsening in both
Zoltan~\cite{devine2006parallel} and KaHyPar~\cite{shhmss2016alenex}, and
explore a range of embedding techniques, including
node2vec~\cite{grover2016node2vec}, MetaPath2Vec++~\cite{dong2017metapath2vec},
FOBE and HOBE~\cite{sybrandt2019heterogeneous}, as well as merged embeddings from
among this set.

\noindent{\bf Node Visit Order.}
We begin matching nodes in an order that tries to prioritize self-similar
regions of the input hypergraph. Specifically, a node is a good candidate for
being contracted at the current level if it shares a hyperedge with a partner
that has a very similar embedding. This indicates that both nodes share many
global structural features that would be preserved in their coarsened
replacement. However, it is also important to reduce the weight of the resulting
coarse nodes. While we also apply more explicit weight-based limitations below,
maintaining the balance of coarse node weights begins with adding a weight
normalization to this embedding-based similarity score.  Otherwise, very dense
regions of the network will be contracted into extremely imbalanced and heavy
nodes before the rest of the hypergraph, which can eventually invalidate the
imbalance constraint.  Note that Section~\ref{sec:prelim} contains more thorough
definitions for the embedding function $\epsilon$ and node weight $w$, as well
as the rest of the notation used in this section. Using these concepts, we can
order nodes based on how similar each is to its closest neighbor. Specifically,
we order each node $u$ with respect to the following:
\begin{equation}
  S_{O}(u) = \max\limits_{v\in\Gamma(u), u \neq v}
             \frac{\epsilon(u)^\intercal\epsilon(v)}{w_uw_v}
  \label{eq:sort_order}
\end{equation}

\noindent{\bf Scoring Contraction Partners.} When visiting node $u$ at a given
level of coarsening, we must select a neighbor $v$ with which it will contract
into a new coarse node in the following level.  To do so, we assign a score to
each neighbor of $u$, and select the node with the highest score to match with.
We assign scores based on a combination of the KaHyPar ``heavy edge'' scoring
function~\cite{hs2017sea}, as summarized in Section~\ref{sec:background}, as
well as the dot product of embeddings.  The heavy edge scoring function
increases the score of hyperedges with fewer nodes. In real-world applications,
this can correspond to ``niche'' communities that tend to carry more meaning for
those involved.  We additionally penalize this score by the node's weights in
order to reduce the imbalance of the resulting coarse nodes.  Specifically, we
assign a score to neighboring nodes $u$ and $v$ during the matching process
equal to:
\begin{equation}
  S_\epsilon(u, v) =
  \left(
  \frac{\epsilon(u)^\intercal\epsilon(v)}
       {w_u w_v}
  \right)
  \left(
  \sum_{e\in\Gamma(u)\cap\Gamma(v)}\frac{w_e}{|e|-1}
  \right)
  \label{eq:score}
\end{equation}

Note that in order for a node pair to receive a high $S_\epsilon$ score, they must both
share low-participation hyperedges as well as global structural embedding-based
features. This way, embedding-based coarsening allows us to break ties between
multiple nodes that all co-occur in similar intersections of similarly weighted
hyperedges, which has the effect of breaking ties, as depicted in
Figure~\ref{fig:part_example}. Additionally, this measure provides a sorting
criteria who's relative values is more important than its absolute value. For
this reason we observe a significant benefit by \emph{not} normalizing the dot
product value. While some embedding techniques encode node similarity through
cosine similarity, which normalizes the dot products between nodes, others do
not. In these cases, the relative magnitudes of embedding dot products is a
valuable signal for determining coarsening partners.

\noindent{\bf Imbalance Constraint.} As previously stated, it is also important
to ensure that the weight of coarsened nodes remains reasonably balanced so that
no coarse node becomes so ``heavy'' that the overall partitioning becomes
imbalanced. To address this, we only match nodes that will produce coarse nodes
below a given weight tolerance. 
We accept the weight tolerance $w^{(T)}$ to be a hyperparameter determined by
the partitioner. Then, when matching nodes $u$ and $v$, we disqualify any pair
such that $w_u + w_v > w^{(T)}$.

\noindent{\bf Embeddings.} Embedding-based coarsening accepts any pretrained embedding ($\epsilon$) to score potential coarsening partners. We assume that this function places each node of the initial hypergraph into a fixed-dimensional space. Because graph embedding is computationally expensive, we interpolate coarse node embeddings from the initial set. This
interpolation consists of the average of all initial node embeddings present in
the coarsened node. For instance, if the coarse node $u$ has weight $w_u$, then
that number of nodes from the input hypergraph have been accumulated into $u$.
These initial nodes, $v_1,\dots,v_{w_u}$ each have embeddings that were supplied
in the initial hypergraph embedding. Therefore, we define $\epsilon(u)$ to be
the following in the case where $u$ is a coarse node that \emph{does not appear}
in the initial embedding:
\begin{equation}
  \epsilon(u)= \frac{1}{w_u}\sum_{i=0}^{w_u} \epsilon(v_i)
\end{equation}

{\bf Runtime Impacts.} Embedding-based coarsening comes with two runtime
increases that are not present in the fast edge-wise coarsening that is
typically used by KaHyPar and Zoltan.  Firstly, one must perform a graph
embedding to learn $\epsilon$. Secondly, at each level of coarsening, we sort
$V$ in accordance to embedding-based signals.  Graph embedding, in general, is
an expensive machine learning operation, requiring significant time and memory
to sample a graph and learn embeddings for each node. However, because the
proposed embedding-based coarsening algorithm is independent of any particular
embedding technique, the specific resources and time needed to produce a graph
embedding are subject to change. There are a few broad patterns that
most embedding methods follow.  Graph embeddings are learned from a set of
samples. These samples can be the edges of the graph itself~\cite{pbg},
observations determined based on first- or second-order
relationships~\cite{sybrandt2019heterogeneous, tang2015line}, or random-walks of
the graph~\cite{dong2017metapath2vec, grover2016node2vec, perozzi2014deepwalk}.
In each case, the observation capturing process is linear with respect to the
size of the graph. Additionally, these observations can often be collected in
parallel. Next, the observations are formulated into batches for a neural
network to learn embeddings. Each observation is viewed once-per-epoch, and
effects the learned weights of a gain model. Therefore, the complexity of
training is equal to the size of the graph times the complexity of performing
back-propagation of a particular model. Embeddings can also be paralleled, both
by GPU acceleration, as well as through multi-node
computation~\cite{recht2011hogwild}. While the embedding process overall is
certainly expensive, the coarsening algorithm proposed in this work only
requires one embedding of the input graph as a prepossessing step. This may not
be feasible for applications that must partition thousands of midsize
hypergraphs daily, but is likely worth it for any application the relies more on
the quality of the resulting partition.

The second difference in runtime comes from the sorting used to prioritize
coarsening partners during each step of the proposed algorithm. At each
iteration, we visit each node to find its most-similar neighbor in terms of node
embedding, and then order nodes by this measure. In contrast, classical
coarsening randomly orders nodes before identifying partners. While this process
introduces the overhead of sorting, we find that removing randomness to
prioritize self-similar hypergraph regions can significantly improve partitioning
quality while decreasing the quality variance. This may save time overall, as
practitioners could reduce the number of random restarts needed to find a high-quality partition.
The entire embedding-based coarsening algorithm is outlined in Procedure~\ref{alg:coarsening}.

\begin{algorithm}
  \caption{Embedding-based Coarsening.}
  \label{alg:coarsening}
  \begin{algorithmic}[1]
    \ENSURE Produces a set of $(u, v)$ pairs to be contracted in the next level
    of coarsening.
    \STATE $M_u \gets \emptyset ~ \forall u \in V$ \COMMENT{$M$ is the matching array.}
    \STATE Sort $u\in V$ in decreasing order by $S_O(u)$. \COMMENT{Eq.~(\ref{eq:sort_order})}
    \FOR{$u \in V$}
      \IF{$M_u = \emptyset$}
        \STATE $p\gets\emptyset$ \COMMENT{$p$ will be matched with $u$.}
        \STATE $s\gets-\infty$ \COMMENT{$s$ is the score associated with $p$.}
        \FOR{$v \in \Gamma(u)$}
          \IF{$v\neq u$ \AND $M_v=\emptyset$ \AND $w_u+w_v<w^{(T)}$}
            \STATE $t\gets S_\epsilon(u, v)$ \COMMENT{Eq.~(\ref{eq:score})}
            \IF{$t>s$}
            \STATE $s \gets t$, $p \gets v$
            \ENDIF
          \ENDIF
        \ENDFOR
        \IF{$p \ne \emptyset$}
          \STATE $M_p \gets u$, $M_u \gets p$ \COMMENT{Match $u$ and $p$.}
        \ENDIF
      \ENDIF
    \ENDFOR
    \STATE Contract nodes according to $M$.
  \end{algorithmic}
\end{algorithm}

\subsection{Experimental Design}
\label{sec:experiment}

We implement embedding-based coarsening in both KaHyPar~\cite{shhmss2016alenex}
and Zoltan~\cite{devine2006parallel}, and compare the result quality against
KaHyPar with community-based coarsening~\cite{hs2017sea}, KaHyPar with
community-based coarsening and flow-based refinement~\cite{heuer2018network},
Zoltan with standard coarsening~\cite{devine2006parallel},
PaToH~\cite{ccatalyurek2011patoh}, and hMetis~\cite{karypis1998hmetis}. For both
KaHyPar and Zoltan with embedding-based coarsening, we compare embeddings
produced by Node2Vec~\cite{grover2016node2vec},
Metapath2Vec++~\cite{dong2017metapath2vec}, FOBE and
HOBE~\cite{sybrandt2019heterogeneous}, as well as a combined FOBE+HOBE
embedding, and a combined Node2Vec, Metapath2Vec++, FOBE, and HOBE embedding.
The combinations are trained using the semi-supervised joint embedding technique
also presented in~\cite{sybrandt2019heterogeneous}, which merges retrained
embeddings through a combination of auto-encoding and link-predictive
objectives.
For the sake of comparison, we choose 100-dimensional embeddings for all cases
and for all hypergraphs.
We selected the FOBE and HOBE combination as this produces a high
quality embedding in prior work~\cite{sybrandt2019heterogeneous}. We then wanted
to explore a new combination with the whole range of considered embeddings.
Additionally, when comparing performance of embedding-based coarsening within
KaHyPar, we compare both with and without flow-based refinement. Overall, we
explore 18 different partitioning settings with embedding-based coarsening, and
five different partitioners with traditional coarsening strategies.

For each of the 23 total partitioner configurations, we explore 96 total
hypergraphs.  Eighty-six of these are supplied by the SuiteSparse Matrix
Collection~\cite{davis2011university}.  These matrices span a range of domains
including social networks, power grids, and linear systems. We interpret each
matrix $\mathcal{H}$ as the incidence matrix of a hypergraph. In doing so, we
consider each row to represent a node, each column to be a hyperedge, and a
nonzero value in $\mathcal{H}_{ij}$ to indicate node $j$ participates in
hyperedge $i$.  We additionally include ten synthetic hypergraphs that were
designed to test the robustness of the coarsening process, extending a similar
approach to generate potentially hard instances from graphs~\cite{safro2015advanced}.  These graphs are a mixture of
graphs that are weakly connected between each other, with less than $1\%$ of
edges connecting different graphs in the mixture.  In multilevel setting, this
can cause the coarsening process to incorrectly contract edges between different
graphs in the mixture, resulting in uneven coarsening, overloaded refinement and
worse quality of the final solution. This structure can be found in many
real-world graphs, including multi-mode networks~\cite{tang2008community} and
logistics multi-stage system networks~\cite{Stock2006}. We introduce additional
complexity by adding additional $<1\%$ random edges (denoted in the online
appendix as ``W/ Noise'').  Full graphs, as well as scripts used to generate
them are available in the online appendix. Summary statistics for each graph are
supplied in the appendix.

For each partitioner and hypergraph combination, we explore both the ``cut'' and
the ``connectivity'' objective, which can influence the initial solution, as
well as some decisions during refinement across the considered
benchmark\footnote{hMetis cannot optimize the ``connectivity'' objective, and is
therefore omitted from that portion of the analysis.}. Additionally, we explore
a number of partitions ($k$) for powers of 2 from 2 to 128. For each
partitioner, objective, and $k$-value combination, we run at least twenty trials
with different random seeds in order to explore the stability of each scenario.
Overall, we compute over 500,000 different experimental trials across our wide
benchmark, and for each trial we record all relevant hyperparameters and quality
results in a database download supplied in our online appendix.

\noindent{\bf Metrics.}
We report a range of
summary statistics for each proposed method. We are primarily concerned with
partitioning performance, as quantified by the value of the considered objective
value at the end of the multilevel paradigm. However, different hypergraphs have
substantially different optimal objective values. Therefore, we report
\emph{improvement} statistics between two considered partitioners, with one
acting as a baseline for the consideration of the other. A value greater than 1
indicates a \emph{reduction} in the considered partitioner when compared to the
baseline across the same hypergraphs.

Formally, if $P$ is a partitioner configuration, including algorithm, embedding
method (if applicable), $k$, and objective function, and $H$ is a hypergraph
then let $P(H)$ be the resulting value of the objective function given $H$ and a
new random seed. Then, let $G$ be a summary statistic, such as mean, min, max,
or standard deviation. We apply $G$ over $\tau$ trials of a given partitioner
with the same input and different random seeds. The improvement of $P$ with
respect to baseline method $P_B$ for a single hypergraph is determined to be:
\begin{equation}
  I(P, P_B, G, H) = \frac{G(P_B(H)_1,\dots,P_B(H)_{\tau})}
                         {G(P(H)_1,\dots,P(H)_{\tau})}
  \label{eq:single_graph_summary}
\end{equation}

Note that the formulation above places the baseline partitioner in the numerator
because an ``improvement'' is quantified as a \emph{decrease} in objective
value. Therefore, if the proposed partitioner $P$ produces consistently
\emph{lower} objective values than $P_B$, then $I$ will be a number greater than
1.

When comparing two partitioners across the entire benchmark of hypergraphs $D$,
we compute the macro-summary. This means that we first apply the summary
statistic $G$ to each hypergraph's trials separately, before averaging the
results together. Formally, the macro-summary is defined as:
\begin{equation}
  \mathcal{I}(P, P_B, G) = \frac{1}{|D|} \sum_{H \in D} I(P, P_B, G, H)
  \label{eq:macro_summary}
\end{equation}

When making these comparisons, we select $P$ and $P_B$ pairs such that both
partitioners are optimizing the same number of partitions using the same
objective. Additionally, we explore summary functions $G$ including mean, min,
max, and standard deviation. While mean indicates average quality, min and max
indicate worse- and best-case performance, while the standard deviation explores
the variance in the resulting partitioners with respect to the random seed.

\subsection{Results}
\label{sec:results}

We present a range of result summaries following the experimental design
discussed above. For a more in-depth look at our results, we present additional
data regarding each hypergraph, and each trial in our online appendix.

To begin our analysis, we summarize the performance improvement of each
embedding-based coarsening implementation compared to its respective baseline.
For instance, we compare KaHyPar with embedding-based coarsening, using FOBE
embeddings, against KaHyPar using community-aware coarsening. We also compare
KaHyPar with flow-based refinement, as well as Zoltan with and without
embedding-based coarsening. These results are summarized using the
macro-improvement statistic $\mathcal{I}$~(Eq.~\ref{eq:macro_summary}), and with
the trial summary statistic $G=mean$. Improvements as a percentage for both
``cut'' and ``connectivity'' objectives for all considered numbers of partitions
($k$) in Table~\ref{tab:summary_results}.

\begin{table*}
\small
  \setlength{\tabcolsep}{4pt}
  
  \begin{subtable}{0.5\textwidth}\centering
  \begin{tabular}{l*{7}{c}}
\# Parts($k$):&2&4&8&16&32&64&128\\
\hline
KaHyPar&8\%&13\%&10\%&6\%&4\%&3\%&1\%\\
KaHyPar(flow)&9\%&11\%&4\%&2\%&3\%&2\%&0\%\\
Zoltan&48\%&28\%&15\%&14\%&9\%&5\%&3\%\\
  \end{tabular}
  \caption{Average connectivity improvement.}
  \end{subtable}%
  \begin{subtable}{0.5\textwidth}\centering
  \begin{tabular}{l*{7}{c}}
\# Parts($k$):&2&4&8&16&32&64&128\\
\hline
KaHyPar&8\%&16\%&9\%&1\%& 3\%&1\%&0\%\\
KaHyPar(flow)&10\%&11\%&3\%&1\%&1\%&1\%&-1\%\\
Zoltan&51\%&45\%&51\%&41\%&31\%&14\%&8\%\\
  \end{tabular}
  \caption{Average cut improvement.}
  \end{subtable}
  
  \caption{
    Improvement for each implementation of embedding-based coarsening
    when compared to its corresponding baseline for both the cut and
    connectivity objectives. Results each use the FOBE
    embedding instance of embedding-based coarsening. Performance numbers
    correspond to $\mathcal{I}$ macro-summaries (Eq.~\ref{eq:macro_summary})
    where $G=\text{mean}$.
  }
  \label{tab:summary_results}
\end{table*}

The most striking result in this small collection of summaries is the inverse
relationship between improvement and $k$. As the number of partitions increases,
the advantage of embedding-based coarsening decreases. This is due to the manner
that we create interpolated embeddings for coarse nodes. As detailed in
Section~\ref{sec:method}, when a newly coarsened node is introduced at a new
level of the coarsening process, it is assigned an embedding equal to the
average of the initial embeddings it contains. This has the effect of
``smoothing'' the embedding space at the coarse level. As a result of this
smoothing, only major variances between nodes will be captured at the point of
initial solution. For instance, if a hypergraph structure has a set number of
key clusters, it is hard for embedding-based coarsening to identify anything
else at the coarsest level. Higher values of $k$, larger than the number of
identified clusters, therefore do not benefit from this technique.

Future work looking creating more useful coarse node representations is likely
to address this problem. However, simple solutions such as embedding coarse
graph instances has significant challenges. For instance, we find that small
problem instances result in poor embedding convergence across all considered
embedding techniques. Therefore we observed in initial trials, re-embedding
coarse graphs dramatically \emph{decreased} result quality. Additionally, graph
embeddings are computationally expensive, and performing any non-constant number
of embeddings is likely to be infeasible for any real-world problem instance.

\begin{figure*}
\begin{subfigure}[t]{0.69\linewidth}\centering
\includegraphics[width=.99\textwidth]{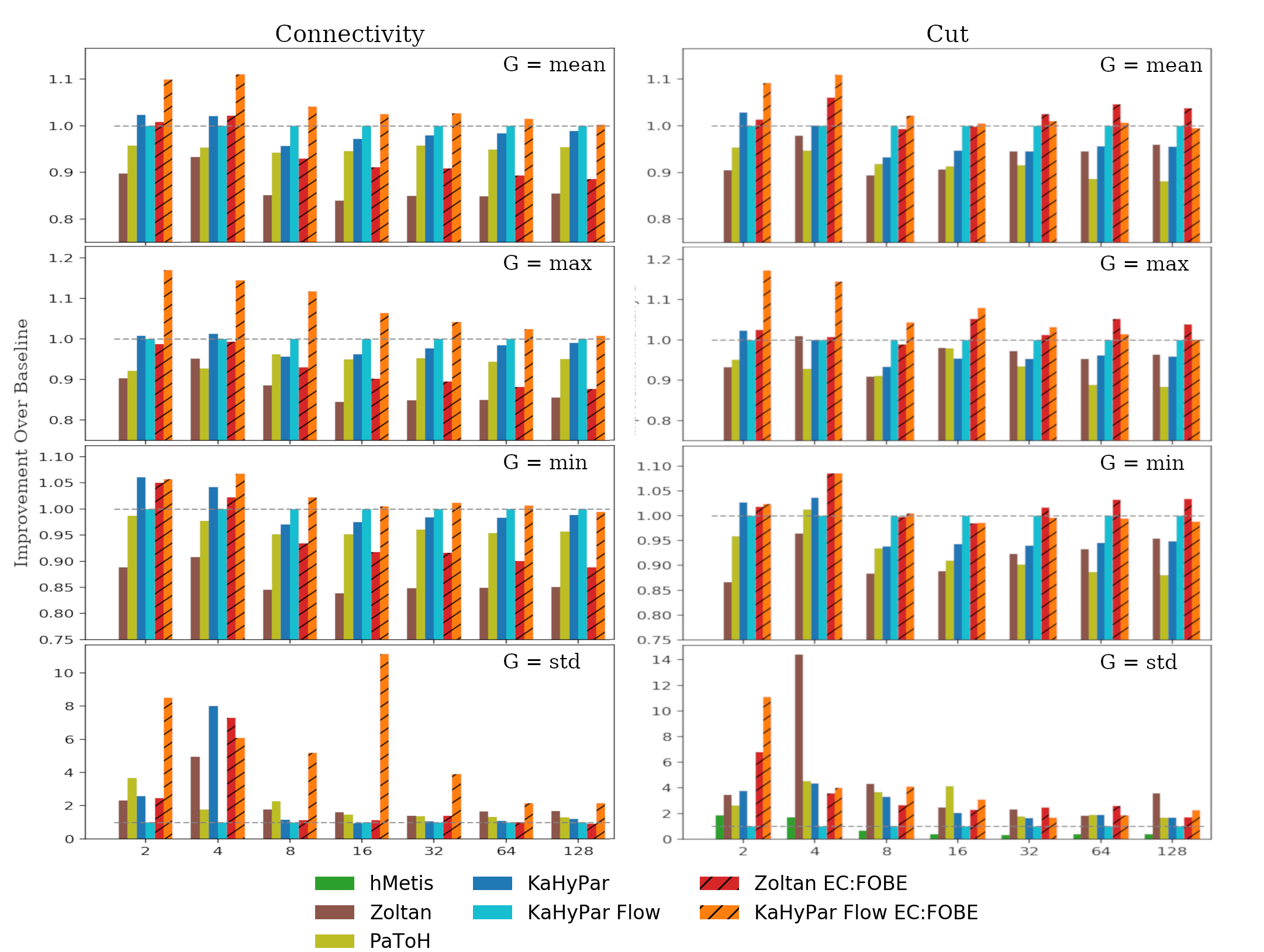}
  \caption{
    The above depicts the relative performance of various partitioners, each
    using KaHyPar with flow-based refinement as a baseline. The results
    correspond to macro-summaries $\mathcal{I}$ (Eq.~\ref{eq:macro_summary}),
    where a value of 1, indicated by the horizontal dashed line, is baseline
    performance of $P_B$. We explore different summary statistics $G$, including
    mean, max, min, and standard deviation.
  }
  \label{fig:baseline_bars}
\end{subfigure}
\hfill
  \begin{subfigure}[t]{0.3\linewidth}\centering
  \includegraphics[width=0.99\textwidth]{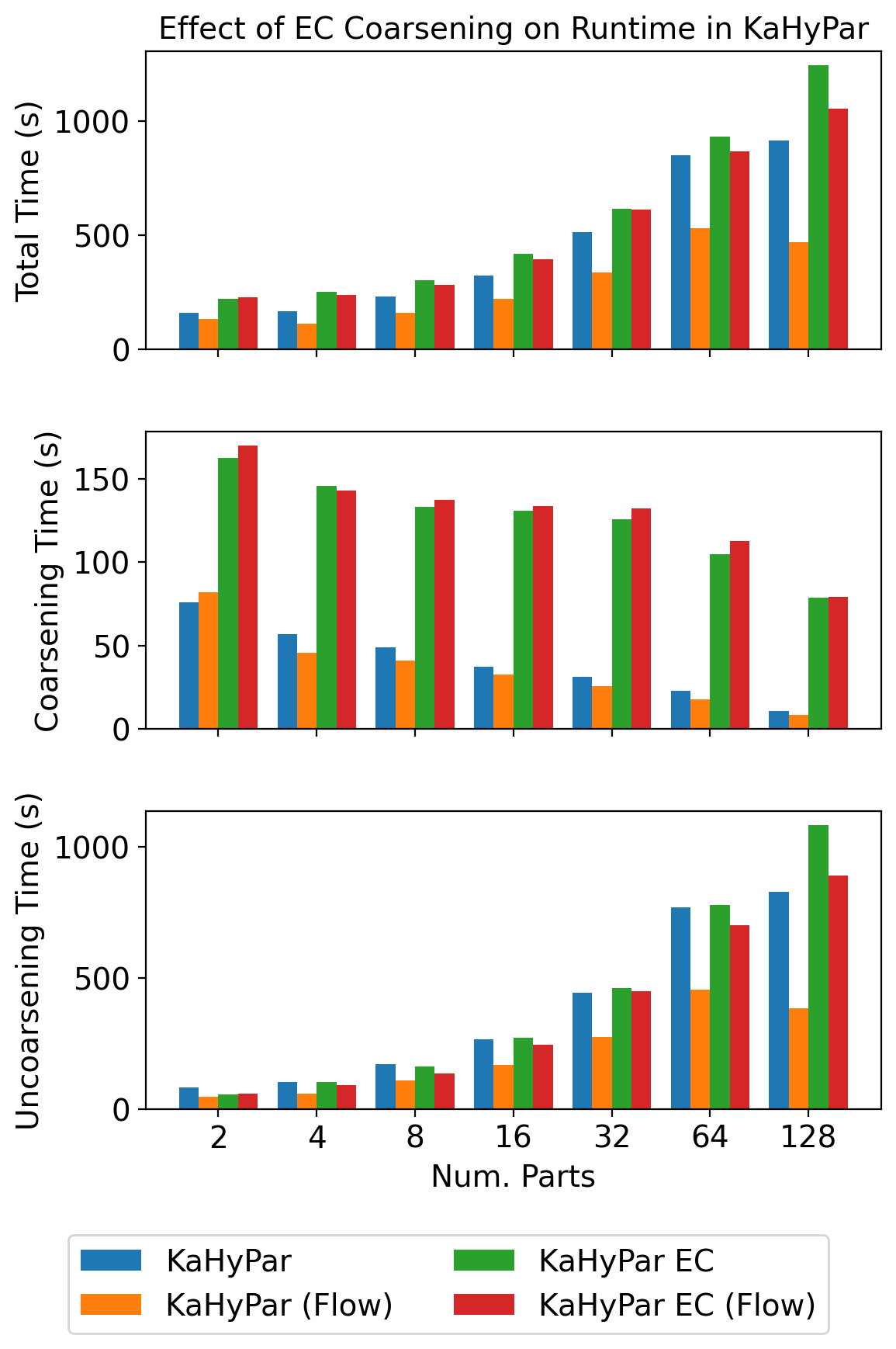}
    \caption{Average Runtime of KaHyPar Across Benchmark.}
    \label{fig:runtime}
  \end{subfigure}
\caption{Effect of Quality and Runtime}
  
\end{figure*}

We continue our comparison of embedding-based coarsening across a range of
baselines in Figure~\ref{fig:baseline_bars}. Here we compare Zoltan with
embedding-based coarsening, KaHyPar with embedding-based coarsening and
flow-based refinement (the better performing KaHyPar implementation), against
all baseline methods. We additionally explore a range of summary statistics $G$
including mean, best-case (min), worse-case (max), and standard deviation. To
easily compare all partitioners, we use KaHyPar with flow-based refinement as
the baseline ($P_B$) for all methods. Therefore, KaHyPar with flow always scores
a one, denoted by the dashed line in each plot, and an improvement over this
baseline is indicated by a macro-improvement $\mathcal{I}$ greater than one.

We observe a similar negative relationship between $k$ and improvement across
the benchmark in Figure~\ref{fig:baseline_bars} as was seen in
Table~\ref{tab:summary_results}. When considering the connectivity objective for
$k$ values of 2 and 4, we interestingly observe that both the KaHyPar and Zoltan
implementations with embedding-based coarsening outperform the baseline. This is
especially important for Zoltan, which greatly under performs the baseline
without our proposed coarsening. When looking at best-base performance
($G=\text{max}$) we observe that the negative trends with respect with $k$ is
less pronounced for KaHyPar and the connectivity objective. This trend
demonstrates the consistent ability of embedding-based coarsening to identify
higher quality solutions than those found by any considered
baseline, and suggests that practitioners willing to accept a quality-for-speed
trade-off can find substantial performance gains with our proposed technique.

Examining the standard deviation results shown in
Figure~\ref{fig:baseline_bars}, we observe that embedding-based coarsening
greatly improves the standard deviation of possible results for a given
hypergraph (shown where $G=\text{std}$). This decrease in variance comes from
the deterministic node-visit order, which replaces a typically random ordering.
As a result, the standard deviation of KaHyPar with embedding-based coarsening
can be reduced by over an order-of-magnitude in some cases. Because many
applications run multiple partitioning trials with various random
seeds in order to find a top-performing result~\cite{trifunovic2006parallel}, we
find that this decrease in variance enables these applications to run fewer
trials while retaining the same confidence in their performance.

\noindent{\bf Comparison with Community-aware Coarsening.} Embedding-based
coarsening attempts to merge together self-similar regions of the input
hypergraph with respect to the structural signals provided by node embeddings.
In contrast, community-aware coarsening restricts the contraction of nodes that
do not share a cluster assignment in the original hypergraph. While these two
approaches are very similar, they both promote contractions within self-similar
regions of the original hypergraph, we find that embedding-based coarsening is a
more flexible constraint. Embedding-based coarsening simply penalizes nodes that
do not share structural features, but may still merge seemingly dissimilar
neighbors if no better options are found. Because of this relaxation, we find
that embedding-based coarsening outperforms community-aware coarsening in a
range of scenarios. This behavior, which we first report in aggregate in
Table~\ref{tab:summary_results}, is depicted in depth in
the appendix. In this example, each considered hypergraph
is listed, and graph-wise performance summaries
$I$~(Eq.~\ref{eq:single_graph_summary}) are depicted for each. The specific
properties of each graph are briefly summarized in the appendix, with more 
information available online.

In the summary table, we demonstrate that for low $k$-values, that
embedding-based coarsening can improve result quality over community-aware
coarsening by around $10\%$ for $k=2,4,8$. When viewing the per-hypergraph
results, we see a more detailed picture. Some hypergraphs with particularly
useful structural features, such as then hypergraph constructed from the enron
email dataset, the eu email dataset, or the difficult and noisy merged
hypergraphs, \emph{can find partitioning solutions with a connectivity objective
that is between one half and one fourth of the community-aware baseline.} For
many other graphs this improvement is a modest few percentage points, while
other graphs are relatively unchanged. For these graphs, we find that the
community-detection solution found by KaHyPar provides nearly the same
information as the selected graph embedding, leading to no improvement.  Only a
small handful of graphs are substantially worsened by this proposed technique
when compared to the community-aware baseline.  For instance, Nemsemm2, a sparse
matrix corresponding to a linear program, is partitioned almost three-times
worse using embedding-based coarsening. The incidence matrix of this hypergraph
is nearly block-diagonal, which results in significant hyperedge-wise features
that are not translated into an embedding, as disjoint graph regions are often
embedded in overlapping spaces. In contrast, Nemswrld is another linear-program
sparse matrix published by the same group, but is less block-diagonal and
receives an statistically significant average improvement of about $33\%$.

\noindent{\bf Comparison Across All Partitioners.}
We supply a large table in the appendix that depicts the average
improvement of each proposed embedding-based coarsening partitioner
configuration against each baseline for the connectivity objective. The numbers
in each cell correspond to the macro-summary $\mathcal{I}$ using $G=\text{mean}$
to summarize trials. For space limitations, we only show this one large table,
but online we present similar tables for the cut objective, as well for the min,
max, and standard-deviation summarizes. In both the in-print and online appendix
we supply improvement numbers per-point of comparison per-hypergraph.
When examining the included table, however, we see clear trends that are
replicated in each online table. 
All considered embeddings improve performance
similarly, with FOBE and HOBE performing marginally above the other methods in
some cases. We observe that Zoltan is the ``easiest'' baseline partitioner,
while KaHyPar with flow-based refinement is the most challenging. Because
KaHyPar with flow-based refinement typically produces higher quality partitions than
Zoltan~\cite{heuer2018network}, it is significant that some
instances of Zoltan with embedding-based coarsening achieve similar
quality.

When looking across the KaHyPar trials, we see that embedding-based
coarsening without flow-based refinement can outperform community-aware
coarsening with the most expensive flow-based refinement. This result confirms
the intuition, initially discussed in Section~\ref{sec:introduction}, that the
coarsening process one of the most fundamental operations in multilevel
partitioning.

Across all considered implementations of embedding-based coarsening we still
observe a decrease in performance for larger values of $k$. As previously
discussed, this derives from the smoothed embedding space produced by iterative
averages of coarse nodes. It is worth noting that this smoothing effect produces
results that are most similar to KaHyPar with its broad community-aware
coarsening. In contrast, embedding-based coarsening still outperforms Zoltan and
PatoH, partitioners that do not account for global structural properties in a
similar way.

\noindent{\bf Runtime}
Our proposed coarsening introduces two sources of overhead into the typical partitioning process, as described in Section~\ref{sec:method}. We compare this runtime effect across our benchmark by focusing on KaHyPar. In Figure~\ref{fig:runtime} we present average runtimes for each implementation, as well as isolated runtimes for the coarsening and uncoarsening phases. Note that when reporting runtime results for the embedding-based implementations of KaHyPar, these performance numbers represent an average across all considered embeddings. We find that embedding-based coarsening multiplies the time needed to coarsen an input hypergraph, which is due to the additional node-wise comparisons and sorting overhead. However, we also find that coarsening amounts to only a fraction of the overall petitioner runtime, which is dominated by the runtime of the uncoarsening phase, which is relatively unchanged.

We plot runtime distributions for FOBE and HOBE, across the benchmark in the supplemental information. These methods are slower and less easily distributed than the other considered embeddings. 
We find that the median considered FOBE embedding requires 1832 seconds, while the median HOBE embedding requires 8682 seconds. 
However, we note that runtime improvements of graph embeddings are an active field of research. Additionally, these runtimes are sensitive to optimizations, hardware, and hyperparameters.
Our graph embeddings add a large factor to the runtime of multilevel partitioning, which
may disqualify our proposed algorithm from ``high performance'' scenarios, such as hypergraph partitioning to accelerate scientific computing workloads at runtime~\cite{hendrickson2006combinatorial}.  However, applications such as placing circuits on a chip~\cite{karypis1999multilevel}, recommending documents~\cite{zhu2016heterogeneous}, machine learning on hypergraphs~\cite{zhou2007learning}, or deep learning on hypergraphs~\cite{feng2019hypergraph}, could all significantly benefit from the slow but higher-quality partitioning brought by embedding-based coarsening.

\section{Conclusion}
\label{sec:conclusion}

We propose embedding-based coarsening, an approach that leverages global
structural features present in a pretrained hypergraph embedding in order
improve the solution quality of multilevel hypergraph partitioning. This
approach prioritizes self-similar regions of the hypergraph by visiting nodes in
a deterministic order based on the embedding properties of each node's
neighborhood. From there, embedding-based coarsening matches nodes by a score
that combines a more traditional edge-wise inner-product with the dot product of
node embeddings. We observe that the introduction of embedding-based features
provides a ``tie-breaking'' mechanism that ultimately preserves global
structural features at the coarsest level in the V-cycle. We implement our
proposed coarsening strategy in both KaHyPar~\cite{shhmss2016alenex} and
Zoltan~\cite{devine2006parallel}.

We observe a significant increase in quality for small values of $k$ (from 2 to
16) gained from embedding-based coarsening. For higher values of $k$ we observe
overall quality that returns to the state-of-the-art baseline.  Furthermore, we
find that embedding-based coarsening improves partitioning quality significantly
across a range of scenarios in both the KaHyPar and Zoltan frameworks.
\emph{Specifically, KaHyPar with flow-based refinement~\cite{heuer2018network}
and embedding-based coarsening, using either FOBE or
HOBE~\cite{sybrandt2019heterogeneous} to produce node embedding, scores
consistently higher on average than all considered baselines.  } Furthermore, we
find that by replacing the random node visit order in many coarsening algorithms
with a deterministic strategy that prioritizes self-similar node pairs, we both
improve solution quality while drastically reducing solution variance, often by
an order of magnitude. Large scale results for all benchmarks and considered
metrics is also available in the online appendix:~\url{sybrandt.com/2019/partition}.

\section{Acknowledgements}

This work was supported by NSF awards MRI \#1725573, DMS \#1522751, and NRT \#1633608.
We would like to thank Sebastian Schlag from the Karlsruhe Institute of Technology for helping us to understand KaHyPar.
We would also like to thank our editor and reviewers.

\bibliographystyle{plain}
\bibliography{main,ruslan}

\pagebreak

\begin{figure*}
  \centering
  \includegraphics[width=.8\textwidth]{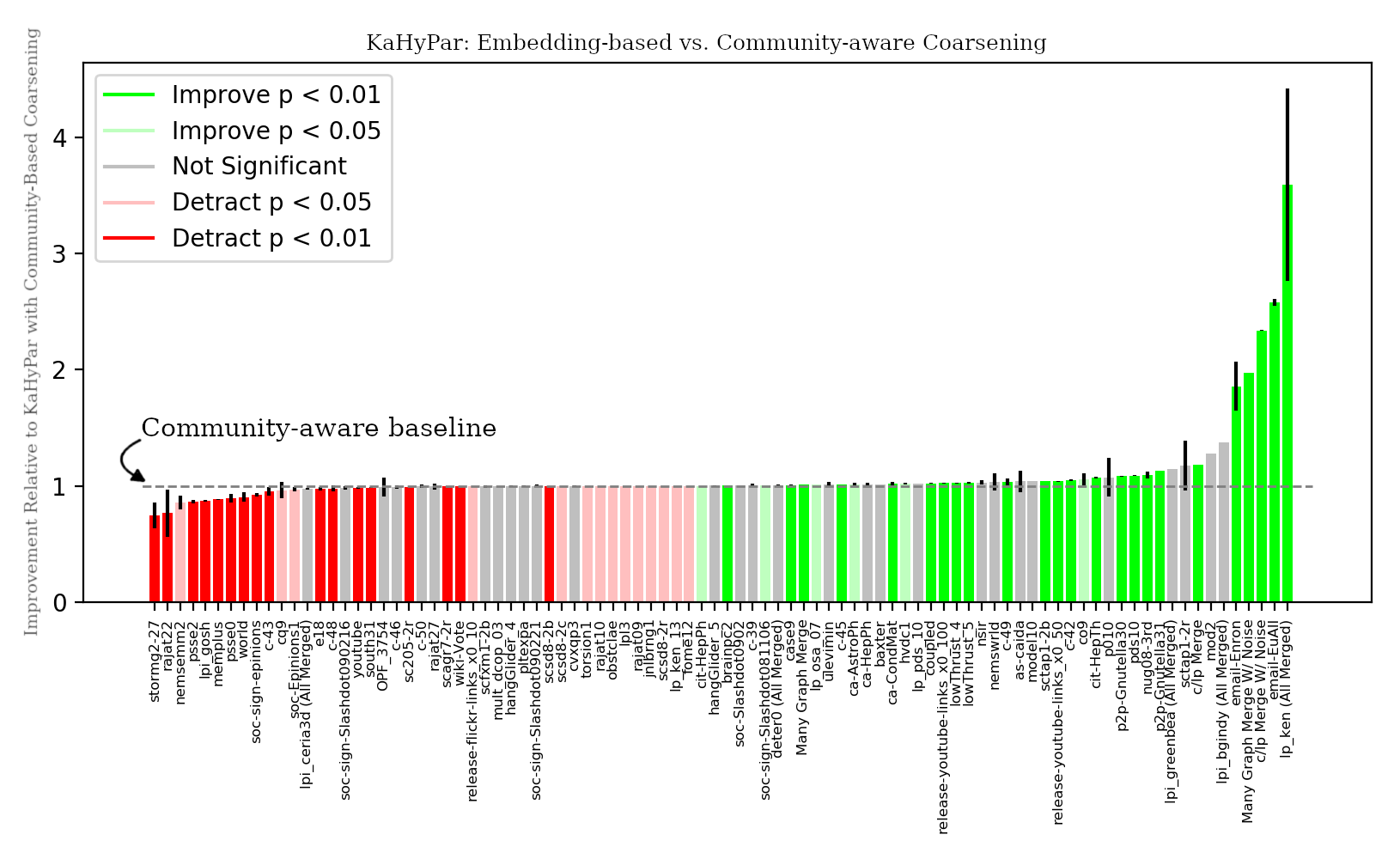}
  \caption{
    The above depicts per-hypergraph summary statistics, $I$ from
    Eq.~\ref{eq:single_graph_summary}, comparing KaHyPar with embedding-based
    coarsening ($P$) to KaHyPar with community-based coarsening ($P_B$). We use
    the mean over trails as our summary statistic $G$, as denoted by the height
    of each bar. A value higher than 1, which is emphasized by the dashed line,
    indicates better solution quality. The small black bar at the top of each
    graph indicates the standard deviation of trials, and the color of each bar
    indicates the statistical significance, where a more saturated color
    indicates a lower $p$-value. Hypergraph names are supplied across the
    horizontal axis, and graphs are ordered by relative improvement.
  }
  \label{fig:kahypar_graphwise_comp}
\end{figure*}

 \begin{sidewaysfigure*}
    \centering
    \includegraphics[width=\textwidth]{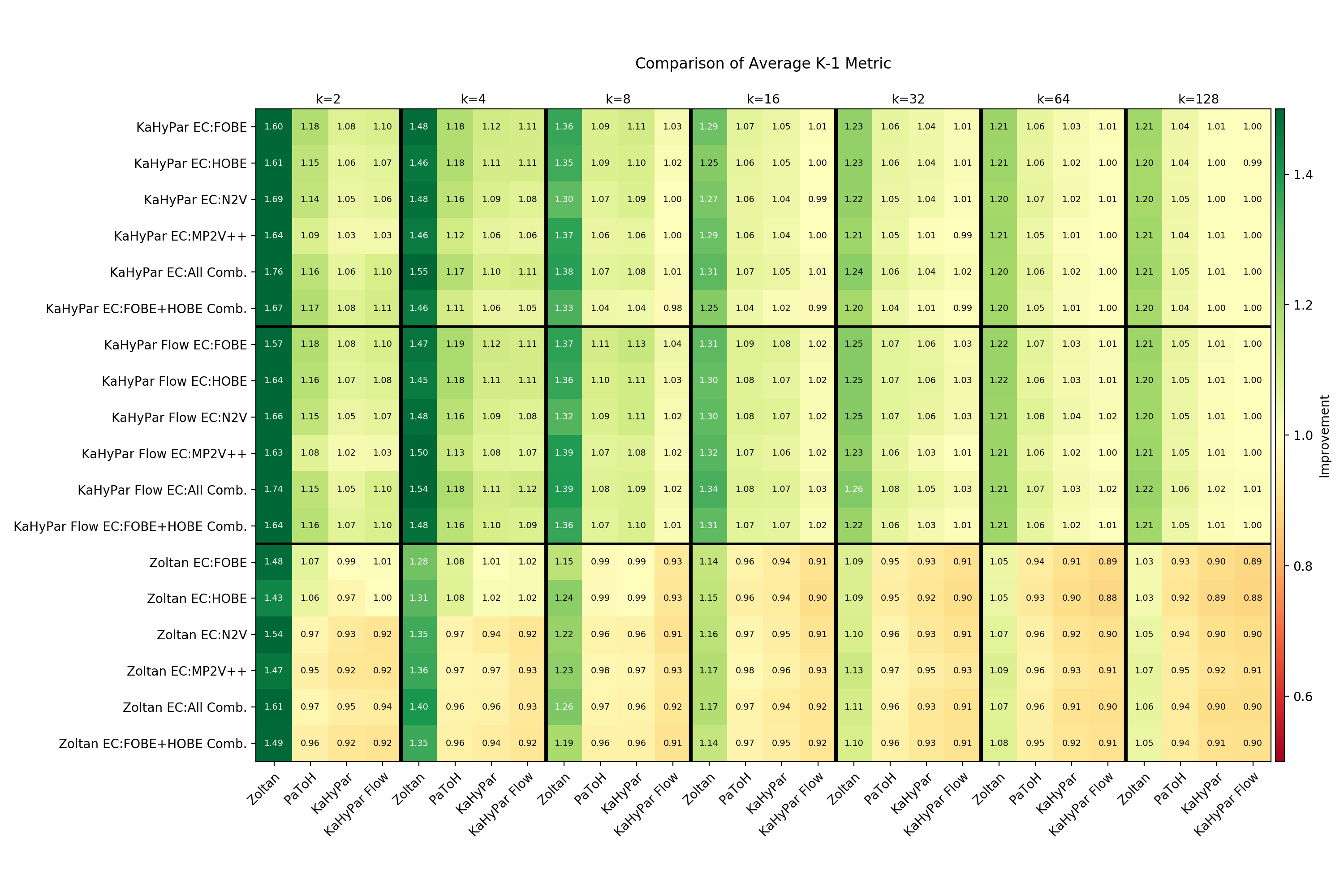}
    \caption{
      The above depicts the average improvement of the connectivity objective
      for all considered partitioners against all baselines. The values in each
      cell correspond to the macro-summary $\mathcal{I}$ using
      $G=\text{mean}$ to summarize trials. The online appendix includes similar
      tables for $G=\text{min, max, and std.}$
    }
    \label{fig:average_km1_matrix}
  \end{sidewaysfigure*}
  
\begin{small}
\renewcommand{\arraystretch}{0.5}
\begin{longtable}{lrrrrrr}
  \caption{
    Hypergraph Details.
    \label{tab:partitioning:hypergraph_details}
  } \\
  $H$ & $|V|$ & $|E|$
  & \multicolumn{2}{c}{$V$ Deg.}
  & \multicolumn{2}{c}{$E$ Size}\\
  & &
  & Mean & Std.
  & Mean & Std. \\
  \hline
  \endfirsthead
  $H$ & $|V|$ & $|E|$
  & \multicolumn{2}{c}{$V$ Deg.}
  & \multicolumn{2}{c}{$E$ Size}\\
  & &
  & Mean & Std.
  & Mean & Std. \\
  \hline
  \endhead
as-caida&26475&16538&3.657&29.016&5.855&29.016 \\
baxter&30722&24255&3.528&5.359&4.469&5.359 \\
brainpc2&27606&27606&6.498&131.257&6.498&131.257 \\
c-39&9271&9271&14.757&41.233&14.757&41.233 \\
c-42&10471&10471&10.532&41.339&10.532&41.339 \\
c-43&11125&11125&11.117&72.602&11.117&72.602 \\
c-45&13206&13206&13.210&85.104&13.210&85.104 \\
c-46&14913&14913&8.744&42.022&8.744&42.022 \\
c-48&18354&18354&9.049&16.866&9.049&16.866 \\
c-49&21132&21132&7.431&14.452&7.431&14.452 \\
c-50&22401&22401&8.644&22.902&8.644&22.902 \\
c/lp Mixture W/ Noise&107776&69568&5.379&12.552&8.333&12.552 \\
c/lp Mixture&107776&69368&5.374&12.552&8.350&12.552 \\
ca-AstroPh&18479&17490&21.369&30.683&22.577&30.683 \\
ca-CondMat&22523&20760&8.194&10.671&8.890&10.671 \\
ca-HepPh&11670&10514&20.181&47.173&22.400&47.173 \\
cari&1200&400&127.333&178.662&382.000&178.662 \\
case9&14453&14453&10.238&105.257&10.238&105.257 \\
cit-HepPh&28093&29526&14.913&27.227&14.189&27.227 \\
cit-HepTh&22908&22610&15.294&43.314&15.496&43.314 \\
co9&22829&10694&4.799&5.264&10.245&5.264 \\
com-dblp-cmty&260998&13477&2.758&4.340&53.411&4.340 \\
coupled&11341&11317&8.685&30.083&8.704&30.083 \\
cq9&21503&9247&4.493&4.673&10.449&4.673 \\
cvxqp3&17500&17500&6.998&3.626&6.998&3.626 \\
deter0 Mixture&21872&7845&2.061&0.900&5.746&0.900 \\
e18&38601&24617&4.053&5.472&6.356&5.472 \\
email-Enron&35153&25481&10.140&33.809&13.989&33.809 \\
email-EuAll&60532&33292&3.765&24.650&6.846&24.650 \\
fome12&48920&24284&2.913&1.303&5.869&1.303 \\
hangGlider\_4&15561&15561&9.609&110.835&9.609&110.835 \\
hangGlider\_5&16011&16011&9.696&112.427&9.696&112.427 \\
hvdc1&24842&24842&6.440&2.936&6.440&2.936 \\
jnlbrng1&40000&40000&4.980&0.141&4.980&0.141 \\
lowThrust\_4&13562&13562&11.867&64.031&11.867&64.031 \\
lowThrust\_5&16262&16262&12.198&70.124&12.198&70.124 \\
lp\_ken Mixture&32418&19219&10.513&10.796&17.733&10.796 \\
lp\_ken\_13&42659&23393&2.157&0.542&3.933&0.542 \\
lp\_osa\_07&25067&1118&5.777&1.032&129.528&1.032 \\
lp\_pds\_10&49932&16239&2.149&0.424&6.607&0.424 \\
lpi\_bgindy Mixture&97920&30322&11.824&9.319&38.182&9.319 \\
lpi\_ceria3d Mixture&39600&35384&11.891&23.733&13.308&23.733 \\
lpi\_gosh&13356&3662&7.474&5.231&27.260&5.231 \\
lpi\_greenbea Mixture&50319&24711&12.288&9.328&25.023&9.328 \\
lpl3&33686&10655&2.979&0.184&9.418&0.184 \\
Graph Mixture W/ Noise&110703&55507&3.650&2.971&7.279&2.971 \\
Graph Mixture&110703&55307&3.645&2.970&7.296&2.970 \\
memplus&17758&17758&7.104&22.035&7.104&22.035 \\
mod2&65990&34355&3.022&2.883&5.804&2.883 \\
model10&16819&4398&8.940&4.645&34.191&4.645 \\
mult\_dcop\_01&25019&24817&7.710&144.682&7.773&144.682 \\
mult\_dcop\_02&25019&24817&7.710&144.682&7.773&144.682 \\
mult\_dcop\_03&25019&24817&7.708&144.682&7.771&144.682 \\
nemsemm2&48857&6922&3.725&2.568&26.292&2.568 \\
nemswrld&28496&6512&6.743&5.108&29.507&5.108 \\
nsir&10055&4450&15.409&25.894&34.817&25.894 \\
nug08-3rd&29856&19728&4.971&3.505&7.523&3.505 \\
obstclae&39996&39996&4.941&0.418&4.941&0.418 \\
OPF\_3754&15435&15435&10.254&5.531&10.254&5.531 \\
p010&19081&10071&6.183&4.984&11.715&4.984 \\
p2p-Gnutella30&36345&9205&2.416&2.594&9.539&2.594 \\
p2p-Gnutella31&62023&15383&2.368&2.669&9.549&2.669 \\
pds10&16558&16558&9.038&7.258&9.038&7.258 \\
pltexpa&70364&26894&2.033&1.288&5.319&1.288 \\
psse0&11028&26694&9.286&6.075&3.836&6.075 \\
psse2&11028&28632&10.452&6.713&4.026&6.713 \\
rajat09&24482&24391&4.309&1.117&4.325&1.117 \\
rajat10&30202&30101&4.311&1.116&4.326&1.116 \\
rajat22&39801&38431&4.919&24.574&5.095&24.574 \\
rajat27&20540&19163&4.786&16.261&5.130&16.261 \\
release-flickr-links\_x0\_10&18612&18612&15.842&38.179&15.842&38.179 \\
release-youtube-links\_x0\_100&115782&115778&3.999&7.222&3.999&7.222 \\
release-youtube-links\_x0\_25&28945&28938&3.996&7.784&3.997&7.784 \\
release-youtube-links\_x0\_50&57891&57888&3.999&7.222&3.999&7.222 \\
sc205-2r&62422&35212&1.974&18.129&3.500&18.129 \\
scagr7-2r&46679&32846&2.574&35.334&3.658&35.334 \\
scfxm1-2b&33047&18266&3.337&6.509&6.038&6.509 \\
scsd8-2b&35910&5130&3.140&17.607&21.982&17.607 \\
scsd8-2c&35910&5130&3.140&17.607&21.982&17.607 \\
scsd8-2r&60550&8650&3.141&22.885&21.990&22.885 \\
sctap1-2b&33858&15390&2.937&17.447&6.462&17.447 \\
sctap1-2r&63426&28830&2.938&23.857&6.464&23.857 \\
soc-Epinions1&50328&31149&9.530&39.646&15.398&39.646 \\
soc-Slashdot0811&77355&70893&11.622&37.228&12.682&37.228 \\
soc-Slashdot0902&82159&71882&11.464&37.486&13.103&37.486 \\
soc-sign-Slashdot081106&64371&27753&7.783&32.163&18.051&32.163 \\
soc-sign-Slashdot090216&68836&30554&7.734&32.971&17.423&32.971 \\
soc-sign-Slashdot090221&69038&30670&7.761&33.115&17.471&33.115 \\
soc-sign-epinions&76359&42470&10.327&43.547&18.567&43.547 \\
south31&35885&17989&3.120&132.364&6.224&132.364 \\
stormg2-27&37485&14306&2.513&2.000&6.584&2.000 \\
torsion1&39996&39996&4.941&0.418&4.941&0.418 \\
ulevimin&46754&6394&3.515&2.714&25.703&2.714 \\
wiki-Vote&2355&3728&43.018&40.735&27.175&40.735 \\
world&66747&34106&2.974&2.751&5.820&2.751 \\
youtube&90581&18173&3.107&8.121&15.487&8.121 \\
\end{longtable}
\end{small}

\end{document}